\documentclass[prd,nofootinbib,onecolumn]{revtex4}
\usepackage{amsmath} %
\usepackage{graphicx} %

\usepackage{units} % Command $\unit[10]{GeV}$ prints correctly 10 GeV. Command
\usepackage[breaklinks=true]{hyperref} %

\IfFileExists{hypernat.sty}{\usepackage{hypernat}}

\usepackage{bm,paralist,xspace}

\begin{document}

\title{Comptonization of cosmic microwave background by cold ultra-relativistic electron-positron pulsar wind
and origin of $\sim$100~GeV lines} %
\author{Dmytro~Iakubovskyi and Sergey Yushchenko}

\affiliation{Bogolyubov Institute for Theoretical Physics, Kiev 03680, Ukraine}

\newcommand{\erf}{\mathop{\rm erf}\nolimits} \newcommand{\tr}{\mathop{\rm
    tr}\nolimits} \newcommand{\sgn}{\mathop{\rm sgn}\nolimits}
\newcommand{\p}{\partial}
\newcommand{\Sp}{\mathop{\rm Sp}\nolimits}
\newcommand{\dm}{{\textsc{dm}}} % Dark Matter
\renewcommand{\deg}{{\textsc{deg}}} % Dark Matter
\newcommand{\fd}{{\textsc{fd}}} % Fermi-Dirac
\newcommand{\dw}{{\textsc{nrp}}} % Dodelson-Widrow
\newcommand{\tg}{{\textsc{tg}}} % Tremaine-Gunn
\renewcommand{\sf}{{\textsc{rp}}} % Shi-Fuller
\newcommand{\mdm}{\ensuremath{m_\textsc{dm}}\xspace} % Dark Matter mass
\newcommand{\lfs}{{\Lambda_\textsc{FS}}} % free-streaming length
\newcommand{\qhd}{{Q}} % Hogan-Dalcanton PSD
\newcommand{\m}{\ensuremath{m}\xspace} % Dodelson-Widrow particle's mass
\newcommand{\numsm}{$\nu$MSM\xspace} % nuMSM
\newcommand{\ev}{\:\mathrm{eV}} % keV
\newcommand{\kev}{\:\mathrm{keV}} % keV
\newcommand{\mev}{\:\mathrm{MeV}} % MeV
\newcommand{\gev}{\:\mathrm{GeV}} % GeV
\newcommand{\pc} {\:\mathrm{pc}} % kpc
\newcommand{\kpc}{\:\mathrm{kpc}} % kpc
\newcommand{\mpc}{\:\mathrm{Mpc}} % Mpc
\newcommand{\gpc}{\:\mathrm{Gpc}} % Gpc
\newcommand{\const}{\mathrm{const}} %
\newcommand{\parfrac}[2]{\left(\frac{#1}{#2}\right)}
\newcommand{\eq}[1]{\begin{equation} #1 \end{equation}}
\newcommand{\ml}[1]{\begin{multline} #1 \end{multline}}
\newcommand{\eV} {\ensuremath{\:\mathrm{eV}}}   %
\newcommand{\keV}{\ensuremath{\:\mathrm{keV}}} %
\newcommand{\MeV}{\ensuremath{\:\mathrm{MeV}}} %
\newcommand{\GeV}{\ensuremath{\:\mathrm{GeV}}} %
\newcommand{\cm}{\:\mathrm{cm}} % cm
\newcommand{\km}{\:\mathrm{km}} % km
\newcommand{\s}{\:\mathrm{sec}} % sec
\newcommand{\ks}{\:\mathrm{ksec}} % ksec
\newcommand{\sr}{\:\mathrm{sr}} % sr
\newcommand{\ph}{\:\mathrm{ph}} % ph
\newcommand{\g}{\:\mathrm{g}} % g
\newcommand{\cts}{\:\mathrm{cts}} % cts
\newcommand{\arcmin}{\:\mathrm{arcmin}} % arcmin
\newcommand{\xmm}{\textsl{XMM-Newton}\xspace} %

\newcommand{\kn}{{\mathrm{dKN}}} % deep Klein-Nishina
\newcommand{\tho}{{\textsc{T}}} % Thomson
\newcommand{\ext}{\mathrm{ext}} % external
\newcommand{\cmb}{{\textsc{cmb}}} % CMB
\newcommand{\mmin}{\mathrm{min}} % minimal

%%%%%%%%%%%%%%%%%%%%%%%%%%%%%%%%%%%%%%%%%%%%%%%%%%%%%%%%%%%%%%%%%%%%%%%

\begin{abstract}
Previously,~\cite{Aharonian:12b} proposed an astrophysical explanation of
narrow gamma-ray line-like feature(s) at $\sim 100$~GeV from Galactic Center region 
observed by Fermi/LAT~\cite{Boyarsky:12}.
The model of~\cite{Aharonian:12b} is based on the inverse Compton scattering of external 
ultra-violet/X-ray radiation 
by a cold ultra-relativistic $e^{+}$-$e^{-}$ pulsar wind. 
We show that the extra broad $\sim$30~MeV component should arise from Comptonization of cosmic microwave 
background radiation. We estimate the main parameters of this component and show that it can be detectable 
with MeV telescopes such as CGRO/COMPTEL. The location of
CGRO/COMPTEL unidentified source GRO~J1823-12 close to excess of 105-120~GeV emission 
(Reg.~1 of~\cite{Boyarsky:12}) can be interpreted as an argument in favour of astrophysical model
of the narrow feature(s) at $\sim$100~GeV.
\end{abstract}

%%%%%%%%%%%%%%%%%%%%%%%%%%%%%%%%%%%%%%%%%%%%%%%%%%%%%%%%%%%%%%%%%%%%%%%

\maketitle

\section{Introduction}

Previous claims about the presence of narrow line-like $\gamma$-ray feature(s) 
around 100-130~GeV observed by Large Area Telescope (LAT) on-board Fermi gamma-ray observatory near the 
Galactic Centre have been received a lot of attention. 
Proposed interpretations include dark matter 
annihilation~\cite{Bringmann:12,Weniger:12,Tempel:12,Su:12,Buchmuller:12},
dark matter decay~\cite{Buchmuller:12,Park:12,Endo:13}, 
systematic effects~\cite{Boyarsky:12,Ackermann:13,Fermi:15} and an astrophysical mechanism -- 
comptonization (in deep Klein-Nishina regime) of cold ultra-relativistic $e^+$-$e^-$ 
pulsar wind by external UV/X-ray emission~\cite{Aharonian:12b}.

In this paper, we discuss the astrophysical mechanism proposed by~\cite{Aharonian:12b} in more details.
We show that, in addition to $\sim$100~GeV line emission, it should produce 
broad $\gamma$-ray component due to comptonization of cosmic microwave background (CMB) radiation. 
For $\sim$100~GeV astrophysical lines, 
the typical energy of this component should be several tens of MeV.
Further detection of unidentified MeV sources with positions coinciding with GeV line excesses
(such as GRO~J1823-12~\cite{Schonfelder:00}, located very close to 105-120~GeV excess 
(Reg.~1 of~\cite{Boyarsky:12})) will argue in favour of the astrophysical origin of GeV lines.

\section{Model description}

The model of~\cite{Aharonian:12b} is based on inverse Compton scattering of energetic (UV and X-ray) photons 
by a cold ultra-relativistic $e^+$-$e^-$ pulsar wind accelerated 
in the vicinity of pulsar magnetosphere, see e.g.~\cite{Aharonian:12} for details.
In this case, the scattering occurs in deep Klein-Nishina regime where the typical energy of the scattered
photon is close to that of initial electron. If conversion efficiency is large enough, 
the mechanism of~\cite{Aharonian:12b} can produce narrow $\sim$100~GeV lines
with flux $F_{line} \sim 10^{-10}~\unit{erg\ cm^{-2}\ s^{-1}}$, consistent with Fermi/LAT observations.
Given the distance to Galactic centre $\sim 8\kpc$~\cite{Malkin:12}, such flux
corresponds to luminosity $\sim 10^{36}~\unit{erg/s}$.

In this paper, we assume the validity of the model proposed in~\cite{Aharonian:12b}. In this case,
one should also detect the softer continuum component due to Compton scattering of CMB radiation on $e^+$-$e^-$ 
wind. In Thomson regime, the average energy of the scattered CMB photons is
\footnote{Throughout this paper, we use notations from~\cite{Blumenthal:70}.}~\cite{Blumenthal:70}
\begin{equation}\langle \epsilon_1 \rangle  = \frac43 \gamma^2 \langle \epsilon_\cmb\rangle \approx
34 \left(\frac{\gamma}{2\times 10^5}\right)^2 \MeV,\label{eq:epsilon1}
\end{equation}
where $\gamma \sim 2\times 10^5$ is the Lorentz factor of cold electron-positron wind (able to produce 
$\sim 100$~GeV photon line~\cite{Aharonian:12b}), 
$\langle \epsilon_\cmb\rangle \approx 2.7\ T_\cmb = 6.3\times 10^{-4}\eV$ is the average energy of CMB photons.
Total flux of this softer component equals to
\begin{equation}
F_{soft} = F_{line}\times\frac{\left(dE/dt\right)_\tho}{\left(dE/dt\right)_\kn}
,\label{eq:Fsoft} 
\end{equation}
where $\left(dE/dt\right)_\tho$ and $\left(dE/dt\right)_\kn$ are the average energy loss rates of a single 
$e^-$ or $e^+$
in Thomson and deep Klein-Nishina regimes, respectively. To calculate $\left(dE/dt\right)_\tho$ and 
$\left(dE/dt\right)_\kn$, we use expressions (2.18) and (2.57) from~\cite{Blumenthal:70}:
\begin{equation}
\left(dE/dt\right)_\tho = -\frac43\sigma_\tho c\gamma^2 \mathcal{E}_\cmb, \label{eq:dEt}
\end{equation}
\begin{equation}
\left(dE/dt\right)_\kn = -\frac{3}{8}\sigma_\tho m_e^2 c^5 \times \int\frac{n_\ext(\epsilon)d\epsilon}{\epsilon}
\left[\ln\left(\frac{4\epsilon\gamma}{m_e c^2}\right)-\frac{11}{6}\right].\label{eq:cooling-rate-KN} 
\end{equation}
Here, $\sigma_\tho$ is the Thomson cross-section, $\mathcal{E}_\cmb \approx 0.26~\unit{eV/cm^3}$ 
is the CMB energy density, $n_\ext(\epsilon)$ is the density distribution of external radiation 
 leading to production of $\sim$100~GeV lines.

According to~\cite{Aharonian:12b}, to explain the smallness of measured width of $\sim$ 100~GeV lines,
the energy of external photons should be high enough,
\eq{\epsilon \gtrsim \epsilon_\mmin = 20 \left(\frac{\gamma}{2\times 10^5}\right)^{-1} \eV,} 
to ensure that the corresponding Compton scattering occurs in the deep Klein-Nishina regime.
\cite{Aharonian:12b} proposes two possible origins of external emission with such an energy:
\begin{itemize}
 \item thermal emission from the surface of the neutron star;
 \item thermal emission from the hot companion star (in case of binary pulsar).
\end{itemize}
In both cases, the density distribution of external radiation can be approximated with that of
rescaled blackbody radiation:
\begin{equation}
 n_\ext(\epsilon) = \frac{15\mathcal{E}_\ext}{\pi^4 T_\ext^4}\frac{\epsilon^2}
{\exp\left(\epsilon/T_\ext\right)-1},\label{eq:next}
\end{equation}
where $\mathcal{E}_\ext$ and $T_\ext$ are the total energy density and temperature of external radiation.
Substituting (\ref{eq:next}) into (\ref{eq:cooling-rate-KN}), we obtain in accordance with expression (2.59) 
of~\cite{Blumenthal:70},
\begin{equation}
\left(dE/dt\right)_\kn =  -\frac{15}{16\pi^2}\sigma_\tho c 
\gamma^2 \mathcal{E}_\ext \mathcal{F}\left(\frac{\gamma T_\ext}{m_e c^2}\right), \label{eq:dEkn}
\end{equation}
where
\[\mathcal{F}(x) = \frac{1}{x^2}\left[\ln(4x)-\frac56- C_E - C_l\right], \ x \gg 1,\]
 $C_E \approx 0.5772$ is the Euler's constant, 
\[C_l = \frac{6}{\pi^2}
\sum\limits_{k=2}^\infty\frac{\ln k}{k^2} \approx 0.5700.\]
Substituting (\ref{eq:dEt}) and (\ref{eq:dEkn}) to (\ref{eq:Fsoft}), we obtain
\begin{equation}
F_{soft} = F_{line}\times \frac{64\pi^2}{45} \frac{\mathcal{E}_\cmb}
{\mathcal{E}_\ext \mathcal{F}\left(\frac{\gamma T_\ext}{m_e c^2}\right)}.\label{eq:Fsoft-final}
\end{equation}
The ranges for function $\mathcal{F}$ can be obtained from numerical estimates of $T_\ext$. For neutron stars,
$T_\ext \lesssim 1~\unit{keV}$, see e.g.~\cite{Pons:06}, 
so $\mathcal{F} \gtrsim 4\times 10^{-5}.$ On the other hand, for small $T_\ext$ function
$\mathcal{F}$ has a maximum, $\mathcal{F} \lesssim 0.0560.$ For these values of $\mathcal{F}$,
we obtain the following relation between $\mathcal{E}_\ext$, $F_{soft}$ and $F_{line}$:
\begin{equation}
65~\unit{eV/cm^3} \lesssim \mathcal{E}_\ext \frac{F_{soft}}{F_{line}} \lesssim 9\times 10^4~\unit{eV/cm^3}.
\label{eq:EFF-final}
\end{equation}

To calculate $F_{soft}$, we use CGRO/COMPTEL observations of GeV line sources in MeV band.
Interestingly, one of the detected CGRO/COMPTEL sources, GRO~J1823-12~\cite{Schonfelder:00} 
(see also 3EG~J1823-1314~\cite{Zhang:04}), is located very close to Reg.~1 of~\cite{Boyarsky:12}, 
where the $\sim 4.7\sigma$ excess at 105-120~GeV has been found~\cite{Boyarsky:12}.
The flux from GRO~J1823-12 in 10-30~MeV band is $(1.0\pm 0.2)\times 
10^{-5}~\unit{photon\ cm^{-2}\ s^{-1}}$~\cite{Schonfelder:00}, the corresponding flux from 3EG~J1823-1314
is $(2.7\pm 0.5)\times 10^{-5}~\unit{photon\ cm^{-2}\ s^{-1}}$~\cite{Zhang:04}, which gives us the estimate 
for $F_{soft}$ detectable with CGRO/COMPTEL near the Galactic Centre region 
\[F_{soft}\simeq \left(10^{-10}-10^{-9}\right)~\unit{erg\ cm^{-2}\ s^{-1}}.\]

To detect softer component from $\sim$ 100~GeV line emitter candidates 
(with $F_{line}\sim 10^{-10}~\unit{erg\ cm^{-2}\ s^{-1}}$) with CGRO/COMPTEL,
the value of $\mathcal{E}_\ext$ should be in the range
\begin{equation}
10~\unit{eV/cm^3} \lesssim \mathcal{E}_\ext \lesssim 10^5~\unit{eV/cm^3}.\label{eq:Eext}
\end{equation}
These values of $\mathcal{E}_\ext$ are expected for Galactic Ridge region~\cite{Ferriere:98,Dogiel:02}.

\section{Conclusions}

We showed that the astrophysical mechanism of $\sim 100$~GeV line production proposed
by~\cite{Aharonian:12b} leads to presence of additional softer broadened component originated from inverse 
Compton 
scattering of CMB radiation by cold ultrarelativistic electron-positron wind. The typical energy of the softer
component should be around $30$~MeV and can it thus be detected by MeV telescopes such as CGRO/COMPTEL. 

Further identification of this component with
instruments operating in MeV band may lead to confirmation 
of astrophysical origin of the $\sim 100$~GeV line(s).

\section*{Acknowledgments}

The authors thank Denis~Malyshev and the anonymous Referee for useful comments.
This work is partially supported by the Program of Fundamental
Research of the Physics and Astronomy Division of the NAS of Ukraine.

%\bibliography{astro,preamble,combined_numsm,dima,cosmomc}

\begin{thebibliography}{20}                                                                                                %

\bibitem{Aharonian:12b} F.~Aharonian, D.~Khangulyan, D.~Malyshev,
Astron. Astrophys. \textbf{547}, A114 (2012).

\bibitem{Boyarsky:12} A.~Boyarsky, D.~Malyshev, O.~Ruchayskiy, Physics of the Dark Universe, \textbf{2}, 
90 (2013). 

\bibitem{Ackermann:13} M.~Ackermann et al. (FERMI-LAT Collaboration), Phys. Rev. D \textbf{88}, 082002 (2013).

\bibitem{Fermi:15} The Fermi-LAT Collaboration, Phys. Rev. D \textbf{91}, 122002 (2015).

\bibitem{Bringmann:12} T.~Bringmann, X.~Huang, A.~Ibarra, S.~Vogl, C.~Weniger,
Journ. Cosmol. Astropart. Phys. \textbf{7}, 54 (2012).

\bibitem{Weniger:12} C.~Weniger, Journ. Cosmol. Astropart. Phys. \textbf{8}, 7 (2012).

\bibitem{Tempel:12} E.~Tempel, A.~Hektor, M.~Raidal, Journ. Cosmol. Astropart. Phys. \textbf{9}, 32 (2012).

\bibitem{Su:12} M.~Su, D.~P.~Finkbeiner, eprint arXiv:1207.7060.

\bibitem{Buchmuller:12} W.~Buchm{\"u}ller, M.~Garny, Journ. Cosmol. Astropart. Phys. \textbf{8}, 35 (2012).

\bibitem{Park:12} J.-C.~Park, S.~C.~Park, Phys. Lett. B \textbf{718}, 1401 (2013).

\bibitem{Endo:13} M.~Endo, K.~Hamaguchi, S.~Pei Liew, K.~Mukaida, K.~Nakayama,
Phys. Lett. B \textbf{721}, 111 (2013).

\bibitem{Schonfelder:00} V.~Sch{\"o}nfelder (COMPTEL Collaboration), Astron. Astrophys. Suppl. \textbf{143}, 
145 (2000).

\bibitem{Aharonian:12} F.~A.~Aharonian, S.~V.~Bogovalov, D.~Khangulyan,
Nature \textbf{482}, 507 (2012).

\bibitem{Malkin:12} Z.~Malkin, e-print arXiv:1202.6128.

\bibitem{Blumenthal:70} G.~R.~Blumenthal, R.~J.~Gould, Rev. Mod. Phys. \textbf{42}, 237 (1970).

\bibitem{Pons:06} J.~A.Pons, B.~Link, J.~A.~Miralles, U.~Geppert, Phys. Rev. Lett. \textbf{98}, 1101 (2007).

\bibitem{Zhang:04} S.~Zhang, W.~Collmar, W.~Hermsen, V.~Sch{\"o}nfelder, Astron. Astrophys. \textbf{421},
983 (2004).

\bibitem{Ferriere:98} K.~Ferri{\`e}re, Astrophys. J. \textbf{497}, 759 (1998).

\bibitem{Dogiel:02} V.~A.~Dogiel, H.~Inoue, K.~Masai, V.~Sch{\"o}nfelder, A.~W.~Strong, Astrophys. J. 
\textbf{581}, 1061 (2002).

\end{thebibliography}

\end{document}